\documentclass[pra,twocolumn,showpacs]{revtex4}
\usepackage{epsfig,amsmath,amssymb,mathtools}
\usepackage[hypertex,linkcolor=red]{hyperref}
\def\eg{{e.~g.} }\def\ie{{i.~e.} }\def\Hg#1{{\rm H^{(#1)}_{gate}}}
\def\<{\langle}\def\>{\rangle}\def\Reals{\mathbb{R}}
\begin{document}
\title{The Quantum Field as a Quantum Computer}
\author{Giacomo Mauro D'Ariano} \affiliation{QUIT Group, Dipartimento di
  Fisica ``A. Volta'', via Bassi 6, I-27100 Pavia, Italy}\affiliation{Istituto
  Nazionale di Fisica Teorica e Nucleare, Sezione di Pavia.}
\begin{abstract} It is supposed that at very small scales a quantum field is an infinite homogeneous
  quantum computer. On a quantum computer the information cannot propagate faster than $c=a/\tau$,
  $a$ and $\tau$ being the minimum space and time distances between gates, respectively. It is shown
  that the information flow satisfies a Dirac equation, with speed $v=\zeta c$ and $\zeta=\zeta(m)$
  mass-dependent. For $a/\tau=c$ the speed of light $\zeta^{-1}$ is a vacuum refraction index
  increasing monotonically from $\zeta^{-1}(0)=1$ to $\zeta^{-1}(M)=\infty$, $M$ being the Planck
  mass for $2a$ the Planck length.
\end{abstract}
\maketitle  
It is interesting to explore the possibility that pure information may underlie all of physics. From
what we know, such information should be made of quantum bits ({\em qubits}), instead of classical
bits. A fundamental problem is then to establish if there is something more than Quantum Theory in a
quantum field. Can we say that a quantum field is just a collection of (infinitely many) quantum
systems, each at every ``space point'' (a Planck cell), unitarily interacting with a bunch of other
systems?  Does the continuum play a fundamental role, or it is only a mathematical idealization? Are
space, time, and all physical observables emergent features of a quantum information processing?

Looking at physics as pure information processing means to consider qubits as primitive entities. In
simple words: qubits are not supported by ``matter'', but matter is made of quantum information
patterns. This is the {\em It from bit} of Wheeler \cite{Wheeler}. At the opposite side of pure
speculation, the new information paradigm has an enormous foundational power, reducing the
fundamental theoretical framework of physics to quantum theory only, and forcing the definition of
each physical quantity to be given in operational terms \cite{CUP,puri}. This is for example the
spirit of the Seth Lloyd's proposal of basing a theory of quantum gravity on a quantum computation
\cite{Lloyd}.  The quantum computational network is just the causal network from which the geometry
of space-time should be derived. The idea of deriving the geometry of space from causal networks is
a program initiated by Rafael Sorkin and collaborators more than two decades ago
\cite{Bombelli-Sorkin_(1987)}. More recently the Lorentz transformations have been explicitly
derived from a causal network with topological homogeneity \cite{DT}, thus showing how relativity
can be regarded as emergent from the quantum computation (a ``visual'' proof of time-dilation and
space-contraction was given in Ref. \cite{DAriano:QCFT}). The main idea is that causality naturally
endows foliations on the causal network \cite{Blute,Hardy}, and the choice of a foliation on a
computational circuit corresponds to synchronize subroutine calls to a global clock in a distributed
computation \cite{Lamport}.

In this paper I will consider an unbounded quantum circuit that is dynamically homogeneous, and, for
simplicity, with the topology of gate connections that can be embedded in two dimensions---the
equivalent of 1+1 dimensions. All the results (apart from maybe anticommuting fields) can be
generalized to more than one space dimension. The dynamical homogeneity of the quantum circuit
represents the equivalent of the physical law, which is supposed to hold everywhere and forever. In
such a way the circuit will incarnate a quantum field theory at some very small scale, e.g. the
Planck scale \cite{Bousso}. We will see that the information flow along the circuit naturally
satisfies a Dirac-like equation. And, as an observable consequence of the unitariety of the
evolution, one has a renormalization of the speed of light, resulting in a vacuum refraction index
which depends on the mass of the field, and which effective stops the flow of information at the
Planck mass.

In the (one-dimensional) quantum computer information can flow only in two directions---right and
left---at the speed $a/\tau$ of one-gate-per-step.  Mathematically we describe the information flows
in the two directions by the two field operators $\phi^+$ and $\phi^-$, for the right and the left
propagation, respectively. In equations one has
\begin{equation}\label{nozigzag}
\widehat\partial_t
\begin{bmatrix}\phi^+\\\phi^-\end{bmatrix}=
\begin{bmatrix}c\widehat\partial_x & 0\\ 0&
  -c\widehat\partial_x\end{bmatrix}\begin{bmatrix}\phi^+\\\phi^-\end{bmatrix}, 
\end{equation}
where $c=a/\tau$ is the speed of the flow over the network, and the hat on the partial derivative
will remind us that they are indeed finite-difference, generally extended to more than one gate (see
the following). If we take the maximal information speed $a/\tau=c$ as a universal constant, then
$c$ must be equal to the speed of light. Now, the only way of slowing-down the information flow is
to have it changing direction repeatedly. A constant average speed corresponds to a simply periodic
change of direction, which is described mathematically by a coupling between $\phi^+$ and $\phi^-$
with an imaginary constant.  Upon denoting by $\omega$ the angular frequency of such periodic change
of direction, we have
\begin{equation}\label{zigzag}
\widehat\partial_t
\begin{bmatrix}\phi^+\\\phi^-\end{bmatrix}=
\begin{bmatrix}c\widehat\partial_x & -i\omega\\ -i\omega &
  -c\widehat\partial_x\end{bmatrix}\begin{bmatrix}\phi^+\\\phi^-\end{bmatrix}. 
\end{equation}
The slowing down of information propagation can be considered as the {\em informational meaning of
  inertial mass}, and $\omega$ represents its value. Notice that Eq.  (\ref{zigzag}) is nothing but
the Dirac equation (without spin), which means that the quantum-information processing corresponding
to pure information transfer simulates a Dirac field---the periodic change of direction being the
Zitterbewegung \cite{Thaller}. It is worth emphasizing that Eq.  (\ref{zigzag}) has been derived
only as a general description of a uniform information transfer, without requiring Lorentz covariance.

The analogy with the Dirac equation leads us to write the coupling constant in terms of the Compton
wavelength $\lambda=c\omega^{-1}=\hbar/(mc)$. This allows us to establish the following relation
between $m$ and $\omega$
\begin{equation}\label{myGod}
m=\frac{\tau^2}{a^2}\hbar\omega=\frac{1}{c^2}\hbar\omega.
\end{equation}
Eq. (\ref{myGod}) provides an informational meaning to the Planck constant $\hbar$ as the conversion
factor between the informational notion of inertial mass in sec${}^{-1}$ and its customary notion in
Kg. Also note that equivalence between the two notions of mass in Eq.  (\ref{myGod}) corresponds to
the Planck quantum expressed as rest energy.

I will now show that the unitariety of the information flows produces a renormalization of $c$ when
introducing the coupling $\omega$, namely the Dirac equation (\ref{myGod}) becomes
\begin{equation}\label{Dirac}
i\widehat\partial_t\phi=(ic\zeta\sigma_3\widehat\partial_x+\omega\sigma_1)\phi,\quad
\phi=\begin{bmatrix}\phi^+\\\phi^-\end{bmatrix}
\end{equation}
where $0<\zeta=\zeta(\omega)\leq 1$ and $\zeta=1$ only for $\omega=0$.

Different from a quantum field, in a quantum computation there is no Hamiltonian, since, in order to
have finite average information speed with $\tau$ non infinitesimal, all the gates must produce a
transformation far from the identity.  We can define a local Hamiltonian matrix in terms of the
discrete time-derivative of the field as 
\begin{equation}\label{Hdef}
\Hg{2n}z:=i\frac{z(n\tau)-z(-n\tau)}{2n\tau}=:i\widehat\partial_t z. 
\end{equation}
We are interested in a field evolution linear in the field, whence we restrict attention to gate
unitary operators $U$ that transform the fields linearly as $z_n(t=1)=Uz_n U^\dag=\sum_kU_{nk}z_k$,
$\{z_k\}$ denoting the field operators involved by the gate. Clearly, the the matrix $\mathbf{U}:=
\|U_{ij}\|$ must be itself unitary, and this will also guarantee preservation of (anti)commutation
relations for the field. By taking the adjoint we get $z_i(t=-1)=\sum_jU_{ij}^\dag z_j$.  By
composing the evolution from many gates we derive the {\em path-sum} rules:
\begin{figure}[h]
\includegraphics[width=.25\textwidth]{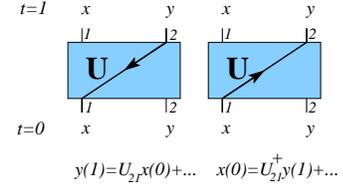}
\caption{Rule for numbering wires to evaluate the contribution of each gate to the forward evolution of 
  the field operator $y$ (left) and to the backward evolution of  $x$ (right).}
\label{f:illus1}
\end{figure}

\begin{figure}[h]
\includegraphics[width=.25\textwidth]{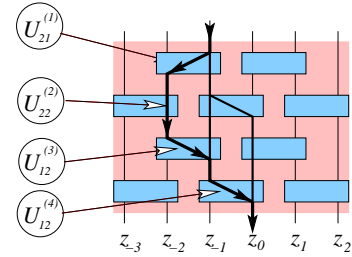}
\caption{Right: illustration of rule for evaluation of a path contribution to the forward evolution
  of the field operator $z_{-1}$ (see text).}
\label{f:illustrule}
\end{figure}

\bigskip\paragraph{Path-sum rule for the forward evolution:}
1) Number all the input wires at each gate, from the leftmost to the rightmost one, and do the
  same for the output wires, as in Figs. \ref{f:illus1} and \ref{f:illustrule}.
2) We say that a wire $l$ is in the past-cone of the wire $k$ if there is a path from $l$ to $k$
  passing through gates.
3) For any output wire $k$ and any input wire $l$ in its causal past cone, consider all paths
  connecting $k$ with $l$, and denote them as follows (see Fig. \ref{f:illustrule})
\begin{equation}
\mathbf i_{kl}=(i_1i_2\ldots i_n i_{n+1})\; \text{with}\; i_1=k,\; i_{n+1}=l,
\end{equation}
4) The following linear expansion holds
\begin{equation}
z_l(t)=\sum_{\mathbf{i}_{kl}} U^{(1)}_{i_1i_2}U^{(2)}_{i_2i_3}\ldots U^{(n)}_{i_ni_{n+1}}z_k(0)
\end{equation}
where $U_{i_mi_{m+1}}$ is the matrix element of the $m$-th gate crossed by the path, from the $i_m$-th
output wire to the $i_{m+1}$-th input wire. 
\paragraph{Rule for evaluating the backward evolution:}
1) For any input wire $l$ and any output wire $k$ in the causal future cone of $l$, consider all
  paths passing through gates connecting $k$ with $l$ (see Fig. \ref{f:illustrule})
\begin{equation}
\mathbf{i}_{lk}=(i_{n+1} i_n\ldots i_2i_1)\; \text{with}\; i_{n+1}=l,\; i_1=k.
\end{equation}
2) The following linear expansion holds
\begin{equation}
z_l(-t)=\sum_{\mathbf{i}_{lk}} U^{(n)\dag}_{i_{n+1}i_n}U^{(n-1)\dag}_{i_ni_{n-1}}\ldots
U^{(1)}_{i_2i_1}z_k(0). 
\end{equation}
\begin{figure}[h]
\includegraphics[width=.25\textwidth]{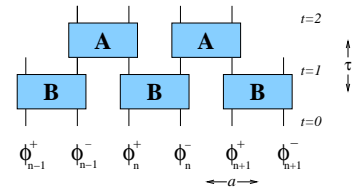}
\caption{Quantum circuit for the Dirac equation (\protect\ref{Dirac}).}\label{f:KG}
\end{figure}
We now derive the field equation corresponding to a quantum circuit describing the interaction
between a left and right-propagating field operators $\phi^\pm$. It is sufficiently general to
consider alternate uniform rows of gates, with unitary interactions $\mathbf{A}$ and $\mathbf{B}$ as
in Fig.  \ref{f:KG}. Using the rules for evolution of the field, one has
\begin{widetext}
\begin{alignat}{2}
&\begin{bmatrix}\phi^+\\\phi^-\end{bmatrix}_{t=2}=&
\begin{bmatrix}A_{21}B_{21}\widehat\partial_-+A_{22}B_{11} &A_{21}B_{22}\widehat\partial_-+A_{22}B_{12}\\
A_{11}B_{21}+A_{12}B_{11}\widehat\partial_+&A_{11}B_{22}+A_{12}B_{12}\widehat\partial_+\end{bmatrix}
&\begin{bmatrix}\phi^+\\\phi^-\end{bmatrix},\\
&\begin{bmatrix}\phi^+\\\phi^-\end{bmatrix}_{t=-2}=&
\begin{bmatrix} B^\dag_{12}A^\dag_{12}\widehat\partial_++B^\dag_{11}A^\dag_{22}& B^\dag_{11}A^\dag_{21}\widehat\partial_-+B^\dag_{12}A^\dag_{11}\\
B^\dag_{21}A^\dag_{22}+B^\dag_{22}A^\dag_{12}\widehat\partial_+&B^\dag_{22}A^\dag_{11}+B^\dag_{21}A^\dag_{21}\delta^-\end{bmatrix}&
\begin{bmatrix}\phi^+\\\phi^-\end{bmatrix}
\end{alignat}
\par\noindent where $\partial_\pm$ denotes the shift operators
$\partial_\pm\vec\phi_n:=\vec\phi_{n\pm 1}$. According to our definition of Hamiltonian in Eq.
(\ref{Hdef}), we have \begin{equation} \Hg4=\tfrac{i}{4\tau}
\begin{bmatrix}A_{21}B_{21}\widehat\partial_--B^\dag_{12}A^\dag_{12}\widehat\partial_++A_{22}B_{11}-B^\dag_{11}A^\dag_{22} &(A_{21}B_{22}-B^\dag_{11}A^\dag_{21})\widehat\partial_-+A_{22}B_{12}-B^\dag_{12}A^\dag_{11}\\
(A_{12}B_{11}-B^\dag_{22}A^\dag_{12})\widehat\partial_++A_{11}B_{21}-B^\dag_{21}A^\dag_{22}&A_{12}B_{12}\widehat\partial_+-B^\dag_{21}A^\dag_{21}\widehat\partial_-+A_{11}B_{22}-B^\dag_{22}A^\dag_{11}\end{bmatrix}.
\end{equation}
It is easy to check that the Hamiltonian is Hermitian, \eg $
\<\phi_n^\pm|\Hg4|\phi_n^\pm\>=\<\phi_n^\pm|\Hg4|\phi_n^\pm\>^*$,
$\<\phi_{n+1}^\pm|\Hg4|\phi_n^\pm\>=\<\phi_n^\pm|\Hg4|\phi_{n+1}^\pm\>^*$, etc. 
\end{widetext} 
In the following we will denote the coarse-grained discrete space-derivative as
$\widehat\partial_x=\frac{1}{4a}(\widehat\partial_+-\widehat\partial_-)$ ($a$ distance between centers of n.n. gates:
see Fig. \ref{f:KG}).
The Hamiltonian $\Hg4$ has the Dirac form (\ref{Dirac}) if
\begin{equation}
\Hg4=c(\mathbf{H}+i\mathbf{K}\widehat\partial_x),
\end{equation}
where
\begin{equation}
\mathbf{H}:=\begin{bmatrix} H_{11} & H_{12}\\H_{12}^* & H_{22}\end{bmatrix},\;
\mathbf{K}:=\begin{bmatrix} K_{11} & K_{12}\\-K_{12}^* & K_{22}\end{bmatrix},
\end{equation}
and
\begin{equation}
\begin{split}
H_{11}=&-\tfrac{1}{2a}\Im(A_{21}B_{21}+A_{22}B_{11})=0,\\
H_{12}=&\tfrac{i}{4a}(A_{21}B_{22}-A^*_{12}B^*_{11}+A_{22}B_{12}-A^*_{11}B^*_{21})=\lambda^{-1},\\
H_{22}=&-\tfrac{1}{2a}\Im(A_{12}B_{12}+A_{11}B_{22})=0,\\
K_{11}=&-\Re(A_{21}B_{21})=\zeta,\\
K_{22}=&\Re(A_{12}B_{12})=-\zeta,\\
K_{12}=&-\tfrac{1}{2}(A_{21}B_{22}-A^*_{12}B^*_{11})=0.
\end{split}
\end{equation}
namely
\begin{equation}\label{ident1}
\begin{split}
&A_{21}B_{22}=A^*_{12}B^*_{11},\\
&A_{21}B_{21}+A_{22}B_{11},A_{12}B_{12}+A_{11}B_{22}\in\Reals,\\
&\Re(A_{21}B_{21})=\Re(A_{12}B_{12})=-\zeta,\\
&A_{22}B_{12}-A^*_{11}B^*_{21}=-4ia\lambda^{-1}.
\end{split}
\end{equation}
Unitarity of $\mathbf{A}$ and $\mathbf{B}$ means
\begin{equation}
\begin{split}
&|A_{11}|^2+|A_{12}|^2=|A_{21}|^2+|A_{22}|^2=1,\\
&A_{11}A^*_{21}+A_{12}A^*_{22}=
A_{21}A^*_{11}+A_{22}A^*_{12}=0,
\end{split}
\end{equation}
and similarly for $\mathbf{B}$. Without loss of generality, we can take the determinants
$|\mathbf{A}|=|\mathbf{B}|=1$, corresponding to $A_{11}=A_{22}^*$, $A_{12}=-A_{21}^*$, and similarly
for $\mathbf{B}$. The first of identities (\ref{ident1}) then gives $B_{11}=B_{22}=0$, whence
\begin{equation}\label{uffa}
\begin{split}
&B_{11}=B_{22}=0,\quad
A_{12}B_{12}=A_{21}B_{21}=-\zeta,\\
&A_{22}B_{12}=-A_{11}^*B_{21}^*=-2ia\lambda^{-1}.\\
\end{split}
\end{equation}
Upon parametrizing $\mathbf{A}$ and $\mathbf{B}$ as follows
\begin{equation}
\mathbf{A}=
\begin{bmatrix}e^{i\phi}\cos\theta &e^{i\psi}\sin\theta \\-e^{-i\psi}\sin\theta &e^{-i\phi}\cos\theta\end{bmatrix},
\;\mathbf{B}=
\begin{bmatrix}0 &e^{i\xi} \\-e^{-i\xi} &0\end{bmatrix},
\end{equation}
one obtains
\begin{equation}
e^{i(\psi+\xi)}=-1,\quad e^{i(\phi-\xi)}=i,
\end{equation}
and
\begin{equation}\label{uffa1}
\sin\theta=\zeta=\sqrt{1-\left(\frac{2a}{\lambda}\right)^2}.
\end{equation}
Eq. (\ref{uffa1}) corresponds to a mass-dependent vacuum refraction index $\zeta^{-1}$ which is
strictly greater than 1 (apart from the special case of zero mass), and monotonically increasing
versus the mass and infinite (i.e. no propagation of information) for $\lambda\to 2a$. Notice also that
for $m=0$ both unitaries become swaps, modulo a phase.

The existence of a vacuum refraction index is a general feature of the discreteness of quantum
information processing, and comes from imposing that the maximum information speed $a/\tau$ cannot
be greater than the speed of light. The refraction index is simply a consequence of unitarity and
linearity in the field operators, independently on the details of the circuit.  Indeed, an upper
bound for $\zeta$ holding for any circuit can be established as follows.  In order to obtain Eq.
(\ref{Dirac}) we need a gate Hamiltonian $\Hg{2n}=
ic\zeta\sigma_3\widehat\partial_x+\omega\sigma_1$. The Hamiltonian is Hermitian, whence $U_f=U_b$.
Moreover, we must have the same number $n$ of time-steps $\tau$ and of space-steps $a$, and from the
form of the Hamiltonian we get $n=2$. We thus have
\begin{equation}
\Hg{4}=\frac{i}{4\tau}(U_f-U_f^\dag )
\end{equation}
and taking the norm of both sides we obtain
\begin{equation}
|\!|\Hg{4}|\!|\leq \frac{1}{2\tau}
\end{equation}
The norm is obtained by Fourier transform at wave-vector $k=\pi/(4a)$, giving
\begin{equation}
\frac{\sqrt{\zeta^2+4\tau^2\omega^2}}{2\tau}\leq\frac{1}{2\tau},  
\end{equation}
namely for $\omega=c\lambda^{-1}$ one has $\zeta\leq \sqrt{1-\left(\frac{2a}{\lambda}\right)^2}$.
\begin{figure}[h]\includegraphics[width=.3\textwidth]{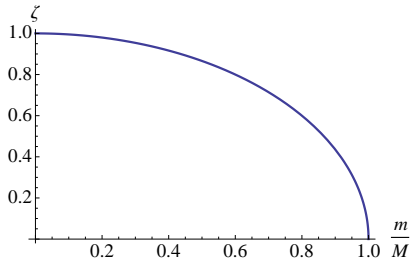}
  \caption{The mass-dependent (inverse) vacuum refraction index $\zeta$ versus the mass $m$ of the
    quantum field. The mass scale is given by $M=\hbar/(2ac)$ ($M$ is the Planck mass for $2a=l_P$
    the Planck length. In the (spin-less) Fermi case this corresponds to two qubits of information
    per Planck length.\label{f:plotbound}}
\end{figure}

Up to now we have considered only an abstract unitary transformation of the field
$\phi=(\phi^+,\phi^-)$. We want now to address the problem of the operator algebras that are
actually processed by the gates. Without loss of generality in the following we will fix the phase
$\phi=0$, whence $\xi=\psi=-\pi/2$. It is easy to show that for field operators $z,z'$ either Bose or
Fermi the following identity holds
\begin{equation}
e^{\alpha^* z'^\dag z-\alpha z^\dag z'} z e^{\alpha z^\dag z'-\alpha^*z'^\dag z} =\cos|\alpha|\,z
+\frac{\alpha}{|\alpha|}\sin|\alpha|\,z'.
\end{equation}
It follows that the unitary operators corresponding to gates $A$ and $B$ have the operator form
\begin{equation}
\begin{split}
A=&\exp\left\{i\theta\left[\phi_n^+{}^\dag\phi_{n-1}^-+\phi_{n-1}^-{}^\dag\phi_n^+\right]\right\},\\
B=&\exp\left\{i\tfrac{\pi}{2}\left[\phi_n^+{}^\dag\phi_n^-+\phi_n^-{}^\dag\phi_n^+\right]\right\},
\end{split}
\end{equation}
where we omit the index $n$ labeling the unitary. The field operators can be written as local
operators in the Bose case, e.~g. $\phi_n^+=a_{2n}$ and $\phi_n^-=a_{2n+1}$, with $a_l$
harmonic-oscillator operators $[a_l,a_k^\dag]=\delta_{lk}$. In the Fermi case we can use the
Clifford algebraic construction
\begin{equation}
\phi_n^+=\sigma_{2n}^-\prod_{k=-\infty}^{2n-1}\sigma_k^3,\quad
\phi_n^-=\sigma_{2n+1}^-\prod_{k=-\infty}^{2n}\sigma_k^3.
\end{equation}
and find 
\begin{equation}
\begin{split}
A=&\exp\left[-i\theta\left(\sigma_{2n-1}^-\sigma_{2n}^++\sigma_{2n-1}^+\sigma_{2n}^-\right)\right],\\
B=&\exp\left[-i\tfrac{\pi}{2}\left(\sigma_{2n}^+\sigma_{2n+1}^-+\sigma_{2n}^-\sigma_{2n+1}^+\right)\right].\\
&
\end{split}
\end{equation}
Therefore, upon associating each wire of the circuit to a local algebra of Pauli matrices, the gate
unitary operators are functions only of the local algebras of their wires. For the vacuum we can
select any state that is left invariant by the quantum computation. In particular, we can choose
$|\mathbf{0}\>=\prod_k|0\>_k$ which is annihilated by the logarithm of either $A$ and $B$, and
similarly for the Bose field. It is easy to see that $N=\sum_n\sigma_n^3$ ($N=\sum_n a_n^\dag a_n$
for Bose) is a constant of motion, which can be interpreted as the number of particles. Notice that
for a given field theory to be simulable by a homogeneous quantum computer in the discrete
approximation $\phi(na)=a^{-\frac{1}{2}}\phi_n$, one needs the field Hamiltonian $H$ (giving
$i\partial_t\phi=[H,\phi]$) that can be written as $H=-\sum_n\phi_n^\dag\Hg{2n}\phi_n$, with $n\geq
1$ satisfying the bound $|\!|\Hg{2n}|\!|\leq \frac{1}{n\tau}$. Such bound gives a general rule for
renormalizing $c$, and with such change all free quantum field theory are simulable.

We conclude by observing that the main results of the present letter hold for space dimension
$d>1$ and upon introducing other degrees of freedom, e.~g. the spin \cite{DUunpub}.

\subsection*{Acknowledgments.} 
I thank Alessandro Tosini and Paolo Perinotti for useful suggestions, and Lucien Hardy, Rafael
Sorkin, and Lee Smolin for very stimulating discussions.


\begin{thebibliography}{15}
\bibitem{Wheeler} J.~A.~Wheeler, in {\em Complexity, Entropy, and the Physics of Information}, ed.
  by W.~Zurek (Addison-Wesley, Redwood City, 1990).
\bibitem{CUP} G.~M.~D'Ariano, in {\em Philosophy of Quantum Information and Entanglement}, ed. by A.
  Bokulich and G.~Jaeger (Cambridge University Press, Cambridge UK 2010).
\bibitem{puri} G. Chiribella, G. M. D'Ariano, and P. Perinotti, Phys. Rev. A {\bf 81} 062348 (2010)
\bibitem{Lloyd} arXiv quant-ph/0501135 (2005).
\bibitem{Bombelli-Sorkin_(1987)} L.~Bombelli, J.~H.~Lee, D.~Meyer, and R.~Sorkin, Phys. Rev. Lett {\bf 59}, 521 (1987).
\bibitem{DT} G. M. D'Ariano and A. Tosini, arXiv 1008.4805 (2010). 
\bibitem{DAriano:QCFT} G.~M.~D'Ariano, in CP1232 {\em Quantum Theory: Reconsideration of
    Foundations, 5} ed. by A.~Y. Khrennikov, (AIP, Melville, New York, 2010), pg. 3.
\bibitem{Blute} R. Blute, I. Ivanov, and P. Panangaden, Int. J. Theor. Phys. {\bf 42} 2025 (2003)
\bibitem{Hardy} L. Hardy, arXiv 0912.4740 (2009)
\bibitem{Lamport} L.~Lamport,  Communications of the ACM {\bf 21} (7), 558-565 (1978).
\bibitem{Bousso} R. Bousso, Rev. Mod. Phys. {\bf 74} 825 (2002).
\bibitem{Thaller} B. Thaller, {\em The Dirac Equation}, (Springer-verlag, Berlin, Heidelberg, New
  York 1992)
\bibitem{DUunpub} G. M. D'Ariano and A. Tosini, unpublished.
\end{thebibliography}
\end{document}